\documentstyle[conf_iap,]{article}
\begin{document}
\heading{%
%
A scaling law of interstellar depletions 
as a tool for \\ abundance studies  
of Damped Ly $\alpha$ Systems\\  
%
} 
\par\medskip\noindent
\author{%
Giovanni Vladilo$^{1}$  
}
\address{%
Osservatorio Astronomico di Trieste, Via Tiepolo 11, I-34131 Trieste, Italy
}

\begin{abstract}
An analytical expression is presented that allows  
dust depletions to be estimated in different types of interstellar environments,
including Damped Ly$\alpha$ systems.  
The expression is a scaling law of a reference depletion pattern
and takes into account the possibility that  the dust 
chemical composition may vary as a function of the dust-to-metals ratio
and of the intrinsic abundances of the medium. 
Preliminary     tests and applications
of the proposed scaling law are  briefly reported. 
\end{abstract}
\section{Introduction}
The presence of dust   can severely affect the
measurement  of   elemental abundances in  Damped Ly $\alpha$ systems (DLAs) 
\cite{PFB,PKSH,HBP}.
Dust depletions in DLAs
resemble those observed in the warm gas of the disk or
halo of the Milky Way \cite{LTWY,KFT}.  
However, disentangling the dust depletion effects from
the intrinsic abundance patterns of DLAs, which  may   differ 
from the solar one, is a difficult task. 
A method for estimating   depletions in DLAs with  minimal assumptions
on their intrinsic abundances   was presented in \cite{V98}. 
In that work the dust in DLAs was assumed to have the same composition
as the dust in Galactic warm gas. The dust-to-gas ratio of individual DLAs 
was then estimated from the observed overabundances of the Zn/Fe ratio,
assuming that the intrinsic  Zn/Fe ratio is solar.   
While successful in estimating dust-corrected abundance ratios in DLAs,
the method \cite{V98} has two limitations:
(1) the dust composition, taken to be constant, may instead vary in different
systems;
(2) the Zn/Fe ratio may   show some deviations
from the solar value \cite{PNC&}.
Here we present
a general scaling law of interstellar depletions that can avoid the above
limitations once  implemented in  
the method \cite{V98}.
A full presentation of this work will be given in a separate paper.

\section{The scaling law}
To derive the scaling law we first define
the fraction in dust of an element X as 
$f_{\rm X} \equiv { N_{\rm {X,dust}} / N_{\rm {X,(gas+dust)}} } $.
By choosing an element Y as a reference for measuring the  dust content
and  the relative abundances, we obtain the expression
$f_{\rm X}  = r  ~ a_{\rm X}^{-1} ~ p_{\rm X}$, where
$r \equiv  f_{\rm Y}$ is the {\em dust-to-metals ratio}, 
 $a_{\rm X} \equiv ({\rm X/Y})_{\rm (gas+dust)}$
 the   relative abundance in the medium,
and  $p_{\rm X} \equiv ({\rm X/Y})_{\rm  dust } $
 the  relative abundance in the dust. 
We then assume that the  chemical composition of the dust is a function of
$r$ and $a_{\rm X}$,
i.e. $p_{\rm X}=p_{\rm X}(r,a_{\rm X})$. 
Therefore $f_{\rm X}=f_{\rm X}(r,a_{\rm X})$ and
from logarithmic  differentiation we obtain  
\begin{equation} 
{ d f_{\rm X} \over f_{\rm X} }
= 
(1+\eta_{\rm X}) ~ { d r \over r } +
(\varepsilon_{\rm X} -1) ~ {d a_{\rm X} \over  a_{\rm X} } ~,
\end{equation}
where 
$ \eta_{\rm X} \equiv  {r \over p_{\rm X}} 
  { \partial p_{\rm X}  \over \partial r }$
and 
$\varepsilon_{\rm X} \equiv  { a_{\rm X} \over p_{\rm X}} ~
 { \partial p_{\rm X}  \over \partial a_{\rm X} }$.  
At this point we integrate Eq. (1) 
assuming $\eta_{\rm X}$ and $\varepsilon_{\rm X}$ to be constant
along the integration path
 running
from  ($f_{{\rm X},i}$, $r_i$, $a_{{\rm X},i}$),
representative of a reference interstellar environment $i$
of solar chemical composition, to 
($f_{{\rm X},j}$, $r_j$,  $a_{{\rm X},j}$),
representative of an arbitrary interstellar environment $j$.
From this integration we derive the scaling law 
\begin{equation}
\label{ScalingLaw}
  f_{{\rm X},j}  
= \left( {r_j\over r_i} \right)^{(1+\eta_{\rm X})}
\, 10^{\left( \varepsilon_{\rm X} - 1 \right) 
\left[ { {\rm X} \over {\rm Y} } \right]_{j} } ~ f_{{\rm X},i} ~,
\end{equation} 
where $[{\rm X/Y}]_j \equiv 
\log ({\rm X/Y})_{j} - \log ({\rm X/Y})_{\circ {\hskip -0.115  cm}
\cdot}$. 
Eq. (\ref{ScalingLaw}) is a generalization of Eq.(11) given in
\cite{V98}. 
The fractions in dust $f_{\rm X}$ are directly related to 
the elemental depletions measured in interstellar studies. 
Therefore the validity of Eq. (\ref{ScalingLaw}) can be  
tested observationally. 
The dependence of $f_{{\rm X},j}$ on $(r_j / r_i)$ can be probed
in the  Milky-Way ISM,  where [X/Y]$_j$ =0
and the depletions are known for many lines of sight.  
From this type of investigation we find  that all the
typical depletion patterns of the Milky Way can be successfully reproduced
with Eq. (\ref{ScalingLaw})
by only varying $(r_j / r_i)$ \cite{V01}. 
To probe the dependence of $f_{{\rm X},j}$ on  $[{\rm X/Y}]_j$
one needs to   measure interstellar depletions in  
   galaxies of known chemical composition, such as the Magellanic Clouds. 
From a study   of   SMC lines of sight  we find evidence
 that these type of observations can indeed be used
to probe and calibrate Eq. (\ref{ScalingLaw}) (work in preparation).  
Once   calibrated with Galactic and extragalactic interstellar observations,
the scaling law (\ref{ScalingLaw})
can be implemented in the method \cite{V98} and applied  
for correcting abundances of DLAs for dust depletion effects.  
The intrinsic Zn/Fe ratio in DLAs can be treated as
a free parameter since deviations of Zn/Fe   from the solar
value are considered in a self-consistent way in Eq. (\ref{ScalingLaw}). 
Preliminary results of application of this revised procedure   
to DLAs indicate that the [Si/Fe]  ratio is, on the average, lower
than in Milky Way stars of similar metallicity \cite{V01}.

\vfill

\begin{iapbib}{99}{
%
\bibitem{HBP} Hou, J.L., Boissier, S., \& Prantzos, N.
2001, A\&A, 370, 23
\bibitem{KFT} Kulkarni, V.P., Fall, S.M., \& Truran, J.W. 1997, \apj, 484, L7
\bibitem{LTWY} Lauroesch, J.T., Truran, J.W., Welty, D.E., \& York, D.G.
1996, PASP, 108, 641
\bibitem{PFB} Pei, Y.C., Fall, S.M., \& Bechtold, J. 1991, \apj, 378, 6
\bibitem{PKSH} Pettini, M., King, D.L., Smith, L.J., \& Hunstead, R.W.
1997, \apj, 478, 536 
\bibitem{PNC&}Prochaska, J.X., Naumov, S.O., Carney, B.W., McWilliam, A., \& Wolfe, A.M.
2000, \aj,  120, 2513
\bibitem{V98} Vladilo, G. 1998, \apj, 598, 534
\bibitem{V01} Vladilo, G. 2001, in 
"Chemical Enrichment of Intracluster and Intergalactic Medium",
Vulcano, May 14-18 2001, in press (astro-ph/0108466)
%
}
\end{iapbib}
\end{document}